\documentclass[
preprintnumbers,
preprint,
a4paper,
showpacs,
amsmath,
amssymb,
byrevtex
]{revtex4}
\usepackage{amsmath,amssymb}
\usepackage{graphicx}
\begin{document}
\title{Static black holes with a negative cosmological constant:
Deformed horizon and anti$-$de Sitter boundaries}
\author{Hirotaka Yoshino}
\email{hyoshino@allegro.phys.nagoya-u.ac.jp}
\author{Tohru Ohba}
\email{tohba@allegro.phys.nagoya-u.ac.jp}
\author{Akira Tomimatsu}
\email{atomi@allegro.phys.nagoya-u.ac.jp}
\affiliation{Department of Physics, Graduate School of Science, Nagoya
University, Chikusa, Nagoya 464-8602, Japan}
\preprint{DPNU-04-04}
\date{\today}
\begin{abstract}

Using perturbative techniques, 
we investigate the existence and properties of 
a new static solution for the
Einstein equation with a negative cosmological constant, 
which we call 
the deformed black hole.
We derive a solution for a static and axisymmetric perturbation of the 
Schwarzschild$-$anti-de Sitter black hole that is regular 
in the range from the horizon to spacelike infinity. 
The key result is that this 
perturbation simultaneously deforms the two boundary surfaces---i.e.,
both the horizon and spacelike two-surface at infinity. 
Then we discuss the Abbott-Deser
mass and the Ashtekar-Magnon one for the deformed black hole,
and according to the Ashtekar-Magnon definition, 
we construct the thermodynamic first law of the
deformed black hole. The first law has a correction term 
which can be interpreted as the work term that is necessary for the
deformation of the boundary surfaces. 
Because the work term is negative, the
horizon area of the deformed black hole becomes larger than that of the 
Schwarzschild$-$anti-de Sitter black hole, if compared under the
same mass, indicating that the quasistatic deformation of the 
Schwarzschild$-$anti-de Sitter black hole may be compatible
with the thermodynamic second law (i.e., the area theorem).    
\end{abstract}
\maketitle
\section{Introduction}

Recently, spacetimes with a negative cosmological constant $\Lambda$
have attracted a lot of attention in various contexts, such as
the AdS/conformal field theory (CFT) correspondence 
(see~\cite{AGMOO00} for a recent review) or 
the Randall-Sundrum brane world scenario~\cite{RS99}. 
One of the necessary investigations in these contexts
would be to analyze the classical feature of the spacetime 
with negative $\Lambda$, 
such as the black hole physics in these spacetimes.

Black hole physics in spacetimes with 
$\Lambda<0$ has a remarkable feature: 
the spatial topology of the black hole horizon is not necessarily spherical
even in stationary four-dimensional spacetimes. 
In the 1990s, new solutions which represent the black holes
with nonspherical topology horizons were  
discovered~\cite{Lem95,Lem95-2,HL95,LZ96,CZ96,ABHP96,Bri96,Man97,KMV98}
in addition to the well-known solutions:  
i.e., the Schwarzschild$-$anti-de Sitter black hole
and the Kerr$-$anti-de Sitter black hole. For example, 
Lemos constructed a static solution of an infinitely large planar black hole
for the four-dimensional spacetime with $\Lambda<0$~\cite{Lem95}. 
He also pointed out that 
the planar solution also represents the cylindrical black hole or
the toruslike black hole by appropriate 
compactifications of this spacetime.
By these procedures, the topology of  
the black hole horizon becomes 
the cylinder ($R\times S^1$)
or the torus ($S^1\times S^1$), and
conformal spacelike infinity has the same topology as the horizon. 
Lemos generalized his static cylindrical black hole solution
to the rotating one~\cite{Lem95-2}.  
The generalization including charge and
dilaton can be found in~\cite{HL95,LZ96,CZ96}. 
There exists another solution of the black hole with unusual topology. 
\AA minneborg {\it et al.} constructed a solution of
a black hole whose horizon is a Riemann two-surface and can 
take an arbitrary genus value $g>1$ by appropriate 
compactifications~\cite{ABHP96}. 
The topology of the conformal spacelike infinity is the same
as that of the horizon due to this compactification. 
The generalization for the charged case can be found in~\cite{Bri96,Man97},
and rotating topological black holes were introduced by
Klemm {\it et al.}~\cite{KMV98}.

Readers might wonder why the black hole solutions with
such various kinds of horizon topology exist even in four dimensions. 
The black hole topology theorem proved by Hawking 
claims that the topology of the horizon  
is ${S}^{2}$~\cite{Hawk72}. But the assumptions of this theorem are
that  the spacetime is 
asymptotically flat and that certain energy conditions hold, which 
are both incompatible with the negative cosmological constant. 
In the presence of negative $\Lambda$, the black hole physics
becomes far richer than the case of $\Lambda=0$. 
On the other hand, there are some theorems that restrict 
black hole solutions in spacetimes with $\Lambda<0$.
Anderson {\it et al.} proved that the Schwarzschild$-$anti-de Sitter
black hole is the unique static solution for asymptotically 
anti$-$de Sitter vacuum spacetimes with $\Lambda<0$~\cite{ACD02}.
Galloway {\it et al.} proved some theorems that
restrict the black hole solutions in $\Lambda<0$ spacetimes~\cite{GSW03}. 
However, we expect that 
there would be a great possibility of the existence of  
unknown solution series.

In this paper, we consider the existence of the series of black hole
solutions which describes a continuous change from 
the Schwarzschild$-$anti-de Sitter black hole
to the cylindrical or planar one. Our expectation is
easily understood by looking at the recent study of black strings 
and black holes in higher-dimensional Kaluza-Klein spacetimes. 
Gregory and Laflamme analyzed the stability of
the black string in higher-dimensional spacetime~\cite{GL93}.
They showed that the black string is unstable for a  
perturbation along the string, if the wavelength is sufficiently large. 
It was also shown that there is a static perturbation, which was a 
strong implication for the existence of a new sequence of static solutions. 
This static perturbation was investigated in detail by Gubser~\cite{Gub02},
and subseqently Wiseman~\cite{Wise03} and Kudoh and Wiseman~\cite{KW03}
numerically solved the sequence that connects the black string to 
the black hole in higher-dimensional Kaluza-Klein spacetime. 
We can expect that a similar situation would occur  
in four-dimensional $\Lambda<0$ spacetimes.

Motivated by this expectation, we analyze the static perturbation of the
Schwarzschild$-$anti-de Sitter spacetime as a starting point.
The existence of such solutions is the manifestation 
of the existence of the new solution series. 
We consider the axisymmetric, even-parity perturbations
in the Regge-Wheeler formalism~\cite{RW57}. 
Each component of the metric perturbation is represented by a product of 
a radial function and Legendre's polynomial $P_l(\cos\theta)$. 
In the case of $\Lambda=0$, the solution diverges either
at the horizon or at infinity~\cite{RW57}. If $\Lambda<0$, however,   
it is possible to construct a solution
which does not diverge everywhere
from the horizon to infinity for all multipole modes 
corresponding to $l=2,3,...$~.  
The horizon of the perturbed solution is not geometrically spherical.
Hence we call this solution a deformed black hole hereafter.  
Because some radial functions
of the perturbation components asymptote to constant values
at spacelike infinity, our solution is not asymptotically anti$-$de Sitter:
a two-surface at spacelike infinity is also deformed. 
This is consistent with the uniqueness theorem of Anderson {\it et al.}
However, our solution still describes the 
weakly asymptotically anti$-$de Sitter spacetime,
in the sense of the Ashtekar-Magnon definition~\cite{AM84}.   
Our result also does not contradict 
the theorems derived by Galloway {\it et al.}

To understand the physical implication of the black hole deformation,
we would like to discuss some features of this solution such as the
mass, the horizon area, and the first law of black hole thermodynamics. 
Because some of the metric coefficients diverge at infinity and
to take a limit to infinity is a delicate problem, 
there are ambiguities in defining the mass of the spacetime with $\Lambda<0$.
Several mass definitions in the asymptotically anti$-$de Sitter spacetimes
have been proposed (see \cite{CN02} and references therein), 
and some of them are applicable also to the
weakly asymptotically anti$-$de Sitter spacetimes. 
Among these, we use 
two well-known mass definitions 
proposed by Abbott and Deser~\cite{AD82} and by
Ashtekar and Magnon~\cite{AM84}, which can be easily applied to 
deformed black holes. 
The first-order perturbation deforms 
the Schwarzschild$-$anti-de Sitter black hole without changing 
the two masses and the horizon area. Hence we should consider the $l=0$ mode 
of the second-order perturbation, 
which is generated by the terms of a product of two first-order perturbation 
components in the second-order equation. 
The two definitions give totally different results  
for the mass derived from 
the second-order perturbation. The Abbott-Deser mass diverges 
to minus infinity, while the 
Ashtekar-Magnon mass gives a finite value. As we will discuss in detail later,
this result would be due to the fact that the Abbott-Deser mass is not 
gauge invariant at second order. Since the definition 
of the Ashtekar-Magnon mass is covariant, 
it would provide a real amount of energy contained in the spacetime
with a deformed black hole. 
Therefore we expect that quasistatic deformation of the
Schwarzschild$-$anti-de Sitter black hole occurs with a finite change 
in the total energy
and thus an investigation of the thermodynamic first law
of the deformed black holes with this mass definition
is meaningful.

One can easily find that
the Schwarzschild$-$anti-de Sitter black holes obey the thermodynamic laws
like asymptotically flat black holes. 
The first law was extended to
the static black holes with unusual topology by Vanzo~\cite{Van97}. 
He used the mass defined as the on-shell value of the Hamiltonian,
which gives the same value as the Abbott-Deser mass, and showed that
the first law holds with this mass definition. 
It was explicitly shown that 
the first law with the usual form also holds 
for the Kerr$-$Newman$-$anti-de Sitter black holes~\cite{CCK00}, 
and there are some approaches to the proof of the first
law for the asymptotically anti$-$de Sitter, 
stationary black hole spacetimes 
under more general assumptions~\cite{Sil02,Bar03}.  
We will analyze the first law of the deformed black holes 
as follows. There are two characteristic scales for 
these spacetimes: i.e., the Schwarzschild radius $R_S\equiv 2m$ 
and the anti$-$de Sitter
radius $R_A\equiv \sqrt{-3/\Lambda}$. 
We analytically consider the case that the Schwarzschild radius is much 
smaller than the anti$-$de Sitter radius (i.e., $\alpha\equiv R_S/R_A\ll 1$) 
and the case that the Schwarzschild radius is much larger 
than the anti$-$de Sitter radius (i.e., $\alpha\gg 1$). 
Then we numerically solve the $\alpha\sim 1$ cases. 
We will explicitly construct the solutions for the first-order perturbation 
and calculate the horizon area and 
the Ashtekar-Magnon mass. 
Our calculation shows that the first law in the usual form approximately 
holds for the $\alpha\gg 1$ case, 
and it does not hold for the other cases. 
In other words, the first law has a correction term which can 
be interpreted as the work term necessary for the deformation of
the two boundary surfaces of the spacetime: i.e., 
the horizon and two-surface at spacelike infinity. 
If we compare the horizon area of the Schwarzschild$-$anti-de Sitter black hole
and that of the deformed black hole under the same mass, 
the latter becomes larger. 
Therefore deformation of the Schwarzschild$-$anti-de Sitter
black hole will be claimed to be a process compatible with
the usual area law.

The outline of the paper is as follows. 
In Sec. II, we derive the equations for a static, axisymmetric 
perturbation of Schwarzschild$-$anti-de Sitter spacetime. 
We show that there exists a solution for the first-order perturbation
which does not diverge everywhere, and that this spacetime is 
the weakly asymptotically anti$-$de Sitter spacetime.
Then we derive the second-order equation for the $l=0$ mode and the general
formula for the horizon area of the deformed black hole.
In Sec. III, we derive the general formulas of the Abbott-Deser mass and
the Ashtekar-Magnon mass for the deformed black holes. 
We discuss the property of these two definitions 
and show that the Abbott-Deser mass is not gauge invariant.  
In Sec. IV, we explicitly construct the solutions 
for the first-order perturbation in two cases  
$\alpha\ll 1$ and $\alpha\gg 1$. 
Then we analyze the first law of the deformed black holes. 
In Sec. V, we numerically calculate for the $\alpha\sim 1$ case and
discuss the dependence of the first law on the value of $\alpha$.  
In Sec. VI, we summarize our results and 
discuss their physical implications.

\section{Static, axisymmetric perturbation}

In this section, we derive the static, axisymmetric perturbation 
of the Schwarzschild$-$anti-de Sitter black hole. 
The background metric is given in the Schwartzschild-like 
coordinates $(t, r, \theta, \phi)$ as follows:
\begin{equation}
d\hat{s}^2=-e^{2\nu_0}dt^2+e^{2\mu_0}dr^2+r^2
\left(d\theta^2+\sin^2\theta d\phi^2\right),
\label{SAD1}
\end{equation}
\begin{equation}
e^{2\nu_0}=e^{-2\mu_0}= 1 - \frac{2m}{r} - \frac{1}{3} \Lambda r^{2}. 
\label{SAD2}
\end{equation}
As we see from the investigations in \cite{RW57} and \cite{GNPP00}, 
the metric of a spherically symmetric spacetime 
with first- and second-order even-parity static perturbations 
can be written in diagonal form 
\begin{equation}
d\hat{s}^2=-e^{2\nu}dt^2+e^{2\mu}dr^2+e^{2\psi}r^2
\left(d\theta^2+\sin^2\theta d\phi^2\right), 
\label{metric_deformed}
\end{equation}
where $\nu$, $\mu$, and $\psi$ are expanded with a small deformation 
parameter $\epsilon$ as follows: 
\begin{align}
&\nu=\nu_0+\epsilon \nu_1+\epsilon^2 \nu_2,
\label{expand_nu}\\
&\mu=\mu_0+\epsilon \mu_1+\epsilon^2 \mu_2,
\label{expand_mu}\\
&\psi=\epsilon \psi_1+\epsilon^2 \psi_2. 
\label{expand_psi}
\end{align}
Here all 
first- and second- order functions are given by
the sum of the products of a radial function
and Legendre's polynomial $P_l(\cos\theta)$, after the
appropriate gauge transformations. 
We use this Regge-Wheeler-like gauge 
for an investigation of the deformed black holes.

\subsection{First-order perturbation}

We give the first-order perturbation as
\begin{align}
&\nu_1=-\mu_1=-H^{(1)}(r) P_{l}(\cos\theta), 
\label{function_nu1_mu1}\\
&\psi_1= K^{(1)}(r) P_{l}(\cos\theta),
\label{function_psi1}
\end{align}
where the first equality in Eq.~\eqref{function_nu1_mu1}
is imposed by the difference of $\theta\theta$, $\phi\phi$ components of
the Einstein equation $R_{\mu\nu}=\Lambda g_{\mu\nu}$. 
We restrict our attention to $l\ge 2$, because
$l=0$ and $l=1$ modes are absorbed to the coordinate transformation and
the changing of the mass. 
The first-order equations become 
\begin{equation}
 r^2e^{2\nu_0}H^{(1)}_{,rr}
+ 2r(re^{2\nu_0})_{,r} H^{(1)}_{,r}
- r^2(e^{2\nu_0})_{,r}K^{(1)}_{,r} 
- \left[2 \Lambda r^2 + l(l+1)\right] H^{(1)} = 0,
\label{Rtt}
\end{equation}
\begin{multline}
 r^2e^{2\nu_0}\left(H^{(1)}_{,rr}-2K^{(1)}_{,rr}\right)
+ 2r\left( re^{2\nu_0}\right)_{,r} H^{(1)}_{,r} \\
- r^2\left[ (e^{2\nu_0})_{,r}+4r^{-1}e^{2\nu_0}\right] K^{(1)}_{,r}
- \left[2 \Lambda r^2- l(l+1)\right] H^{(1)} = 0,
\label{Rrr}
\end{multline}
\begin{eqnarray}
e^{2\nu_0}\left(H^{(1)}_{,r} - K^{(1)}_{,r}\right)
 + \left(e^{2\nu_0}\right)_{,r}H^{(1)} = 0,   
\label{Rr theta}
\end{eqnarray}
\begin{multline}
 r^2e^{2\nu_0} K^{(1)}_{,rr}
- 2re^{2\nu_0}H^{(1)}_{,r}
+ r^2\left[ (e^{2\nu_0})_{,r}+4r^{-1}e^{2\nu_0}\right] K^{(1)}_{,r} \\
- 2\left( re^{2\nu_0} \right)_{,r} H^{(1)} 
-  (l^2+ l  -2)K^{(1)} = 0,
\label{R phi phi}
\end{multline}
where ${,r}$ denotes the derivative with respect to $r$. 
These equations come from $tt, rr, r\theta$ components and
the sum of the $\theta\theta, \phi\phi$ components of the Einstein equation,
respectively. 
Although there are four equations for two functions $H^{(1)}$ and $K^{(1)}$, 
these equations do not overdetermine $H^{(1)}$ and $K^{(1)}$ because
we can derive Eq.~\eqref{Rtt} and the derivative of Eq.~\eqref{R phi phi}
using Eq.~\eqref{Rrr}, Eq.~\eqref{Rr theta},
and their derivatives.
Eliminating $K^{(1)}$ from Eqs.~\eqref{Rtt} and \eqref{Rr theta},
we obtain the equation for the quantity $M\equiv r^2e^{2\nu_0}H^{(1)} $:
\begin{eqnarray}
M_{,rr} - 2\left(r^{-1}+{\nu_0}_{,r}\right) M_{,r} 
-e^{-2\nu_0}\left[ 2\Lambda +r^{-2}(l^2+l-2)\right] M = 0.
\label{H1}
\end{eqnarray}
Eliminating $K^{(1)}_{,rr}$ and $K^{(1)}_{,r}$ 
from Eqs.~\eqref{Rtt}, \eqref{Rrr}, \eqref{Rr theta}, and \eqref{R phi phi},  
we find that $K^{(1)}$ is expressed in terms of $H^{(1)}$ as
\begin{equation}
K^{(1)} = \frac{1}{(l^2+l-2)} 
\left\{ 
r^2(e^{2\nu_0})_{,r}H^{(1)}_{,r}
+ \left[
r^2(2{\nu_0}_{,r})\left(e^{2\nu_0}\right)_{,r}
+l^2+l-2e^{2\nu_0}
\right]H^{(1)}
\right\} .
\label{K1}
\end{equation}

Now we show the existence of solutions which do not diverge 
in the range from the horizon to spatial infinity. 
The two independent solutions for $H^{(1)}$ of Eq.~\eqref{H1} 
are $(r-r_h)$ and $1/(r-r_h)$ in the neighborhood of the horizon,
while $1/r$ and $1/r^2$ for large $r$. 
Here we have introduced the horizon radius $r_h$, which is the maximum 
solution of $e^{2\nu_0}(r_h)=0$. Thus, if we choose 
$(r-r_h)$ near the horizon, the solution $H^{(1)}$ becomes regular
everywhere because it behaves like $\sim 1/r$ at large $r$. 
The behavior of $H^{(1)}$ at infinity can be written as
\begin{equation}
H^{(1)}=\frac{a_1}{r}+\frac{a_2}{r^2}+\frac{a_3}{r^3}+\cdots,
\label{H1asymptotic}
\end{equation}
where $a_3, a_4, ...$ are expressed with $a_1$ and $a_2$ like
\begin{align}
&a_3=-\frac{3(l^2+l-4)}{2\Lambda}a_1,
\label{a3}\\
&a_4=-\frac{1}{2\Lambda}\left[18ma_1+(l^2+l-6)a_2\right],
\label{a4}
\end{align}
because $H^{(1)}$ contains only two integral constants. 
The ratio of $a_1$ and $a_2$ is determined by imposing 
the regularity at the horizon.
Using Eq.~\eqref{K1}, we see that 
the behavior of $K^{(1)}$ at infinity becomes
\begin{equation}
K^{(1)}=c_0+\frac{c_1}{r}+\frac{c_2}{r^2}+\cdots,
\label{K1asymptotic}
\end{equation}
where the coefficients can be determined using Eq.~\eqref{K1}
as follows:
\begin{align}
& c_0=\frac{2\Lambda a_2}{3(l^2+l-2)},
\label{c0}\\
& c_1=-a_1,
\label{c1}\\
& c_2=0.
\label{c2}
\end{align}
Thus, $K^{(1)}$ is also finite at $r=\infty$.
Similarly, if we write the behavior of $H^{(1)}$ near the horizon as
\begin{equation}
H^{(1)}={\tilde{a}_1}(r-r_h)+\tilde{a}_2{(r-r_h)^2}+\cdots,
\label{H1_horizon}
\end{equation}
the behavior of $K^{(1)}$ near the horizon becomes 
\begin{equation}
K^{(1)}=\tilde{c}_0+\tilde{c}_1{(r-r_h)}+\cdots,
\label{K1_horizon}
\end{equation}
where the coefficients can be determined using Eq.~\eqref{K1}
such as
\begin{align}
& \tilde{c}_0=\frac{4(3m-r_h)\tilde{a}_1}{l^2+l-2}, 
\label{tilde_c0}\\
& \tilde{c}_1=2\tilde{a}_1.
\end{align}
Thus two functions $H^{(1)}$ and $K^{(1)}$ remain finite everywhere 
from the horizon to infinity, and hence 
we have established the existence of the new solution series
of $\Lambda<0$ spacetimes.

Now we examine some properties of this solution. 
The induced metric on a two-surface
$t={\rm const}$ and $r=r_h$ becomes
\begin{equation}
d\hat{s}^2=\left[1+2\epsilon \tilde{c}_0 P_l(\cos\theta)\right]r_h^2
(d\theta^2+\sin^2\theta d\phi^2), \label{metric_horizon}
\end{equation} 
which implies that the geometry of the horizon in (say) the $l=2$ case 
is prolate for $\tilde{c}_0>0$ and oblate for $\tilde{c}_0<0$. 
Thus the horizon geometry deviates from the geometrically spherical 
surface and this solution represents the deformed black hole. 
To see the structure of spacelike infinity, we consider a
conformal transformation $ds^2=\Omega^2 d\hat{s}^2$ where
we choose $\Omega=r^{-1}$. The induced metric of $r={\rm const}$ surface 
at infinity of the conformally transformed spacetime becomes
\begin{equation}
ds^2=\frac13\Lambda dt^2+\left[1+2\epsilon {c}_0 P_l(\cos\theta)\right]
(d\theta^2+\sin^2\theta d\phi^2), \label{eq:infinity}
\end{equation}
which means that the conformal boundary of this spacetime is 
also deformed. This indicates that 
this spacetime is not asymptotically anti$-$de Sitter
in the sense of the Ashtekar-Magnon definition~\cite{AM84}.  
According to their definition, the spacetime is asymptotically anti$-$de Sitter
if the spacetime satisfies the Einstein equation with $\Lambda<0$ and with 
an energy momentum tensor that satisfies an appropriate falloff condition, 
and its conformal boundary $\mathcal{I}$ is topologically $S^2\times R$, and
the conformal group of $\mathcal{I}$ is the anti$-$de Sitter group. 
The last condition is equivalent to that
$\mathcal{I}$ admits a global chart $(t, \theta, \phi)$ such that
the metric on $\mathcal{I}$ is conformally related to the metric
\begin{equation}
ds_0^2=\frac13\Lambda dt^2+d\theta^2+\sin^2\theta d\phi^2.
\end{equation}
This is clearly inconsistent with Eq.~\eqref{eq:infinity}.
However, the spacetime of the deformed black hole is weakly asymptotically
anti$-$de Sitter~\cite{AM84}: 
this is defined by the above conditions except the last
condition. 
The deformed black hole spacetime has a deformed surface
at spacelike infinity, 
and thus the existence of this solution is not contradictory 
to the uniqueness theorem 
of the asymptotically anti$-$de Sitter black hole~\cite{ACD02}.

To solve analytically 
$H^{(1)}$ and $K^{(1)}$ for general $\alpha$ is rather difficult.
We will explicitly construct the analytic solutions 
in two simple cases 
$\alpha\equiv 2m\sqrt{-\Lambda/3}\ll 1$ and $\alpha\gg 1$ in Sec. IV
and the numerical solutions in the $\alpha\sim 1$ cases in Sec. V. 
Here, we show that $H^{(1)}$ and $K^{(1)}$ 
do not change their sign for $r_h<r<\infty$.
Using Eq.~\eqref{Rr theta}, $K^{(1)}$ is given by
\begin{equation}
K^{(1)}=H^{(1)}+\int_{r_h}^{r}{2\nu_0}_{,r} H^{(1)}dr+\tilde{c}_0.
\label{K1_positive}
\end{equation}
We consider the $\tilde{a}_1>0$ case (hence $\tilde{c}_0>0$). 
If $H^{(1)}$ changes its sign from positive to negative at $r=r_c$,
$K^{(1)}(r_c)$ should be positive from Eq.~\eqref{K1_positive} and
$H^{(1)}_{,r}(r_c)$ should be negative. However, this contradicts 
Eq.~\eqref{K1}. Hence $H^{(1)}$ is always positive. 
Using Eq.~\eqref{K1_positive} 
again, we see that $K^{(1)}$ also always takes a positive value. 
The proof for the case $\tilde{a}_1<0$ is similar. 
Thus $H^{(1)}$ and $K^{(1)}$
have the same sign and do not change their sign. 
This means that there is a natural relation 
between the geometry of the horizon and that of 
the conformal boundary of this spacetime. 
Because $\tilde{c}_0$ and $c_0$ have 
the same sign, the conformal boundary
is prolate if the horizon is prolate and vice versa. 
Moreover, using Eq.~\eqref{K1_positive} again, we see that
$|K^{(1)}-H^{(1)}|$ is a monotonically increasing function of $r$.
Because $H^{(1)}(r_h)=H^{(1)}(\infty)=0$, 
we find $|K^{(1)}(r_h)|<|K^{(1)}(\infty)|$ and hence $|\tilde{c}_0|<|c_0|$.
Thus the conformal boundary at infinity 
is more deformed compared to the horizon.

As we see from Eq.~\eqref{metric_horizon}, 
the first-order perturbation does not change the area of the horizon.
It also does not affect the mass of the spacetime,
as we will see in the next section. 
Hence to see the change in the mass and horizon area due to the
deformation of the black hole, 
we should consider the $l=0$ mode of the second-order perturbation.

\subsection{$l=0$ mode of second-order perturbation}

We consider the second-order perturbation 
which is generated by the $l\ge 2$ mode of the first-order perturbation
of the black hole.
The second-order perturbation is given as follows:
\begin{align}
&\nu_2=-{H}_0^{(2)}(r)
 +\sum_{n\neq 0}H_{n}^{(2)}(r)P_{n}(\cos\theta),
\label{function_nu2}\\
&\mu_2=\tilde{H}_0^{(2)}(r)
  +\sum_{n\neq 0}\tilde{H}_{n}^{(2)}(r)P_{n}(\cos\theta),
\label{function_mu2}\\
&\psi_2=K_0^{(2)}(r)
  +\sum_{n\neq 0}K_{n}^{(2)}(r)P_{n}(\cos\theta).
\label{function_psi2}
\end{align}
Because there remains the degree of freedom to choose the $r$ coordinate, 
we can impose $H^{(2)}\equiv H_0^{(2)}=\tilde{H}_0^{(2)}$ only 
for the $l=0$ mode. 
We will treat only the functions of the $l=0$ mode,
$H^{(2)}$  and $K^{(2)}\equiv K_0^{(2)}$, for 
the second-order perturbation. 

The second-order Einstein equations of the $l=0$ mode are
\begin{multline}
 r^2e^{2\nu_0}H^{(2)}_{,rr}
+ 2r(re^{2\nu_0})_{,r}H^{(2)}_{,r}
- r^2(e^{2\nu_0})_{,r}K^{(2)}_{,r}
- 2 \Lambda r^2 H^{(2)} \\
= \frac{2}{2l +1} 
\left\{
e^{2\nu_0}r^2
\left[ \left( H^{(1)}_{,r} \right)^{2} - H^{(1)}_{,r} K^{(1)}_{,r}\right] 
+ \left[  \Lambda r^2+  l \left( l + 1 \right) \right] 
    \left( H^{(1)} \right)^{2} 
- l \left( l + 1 \right) H^{(1)} K^{(1)} 
\right\},
\label{R2tt}
\end{multline}
\begin{multline}
 r^2e^{2\nu_0}\left(H^{(2)}_{,rr}-2K^{(2)}_{,rr}\right)
+ 2r(re^{2\nu_0})_{,r} H^{(2)}_{,r} 
-r^2\left[(e^{2\nu_0})_{,r}+4r^{-1}e^{2\nu_0}\right]K^{(2)}_{,r}
- 2 \Lambda r^2 H^{(2)} \\
= \frac{2}{2l +1} 
\Big\{
e^{2\nu_0}r^2
\left[ \left( H^{(1)}_{,r} \right)^{2} 
  + \left( K^{(1)}_{,r} \right)^{2}
  - H^{(1)}_{,r} K^{(1)}_{,r} \right] 
+ \left[   \Lambda r^2  -  l \left( l + 1 \right) \right]
  \left( H^{(1)} \right)^{2}\\
+   l \left( l + 1 \right) H^{(1)} K^{(1)} 
\Big\},
\label{R2rr}
\end{multline}
\begin{multline}
  r^2e^{2\nu_0}K^{(2)}_{,rr}
- 2re^{2\nu_0}H^{(2)}_{,r}
+ r^2\left[ (e^{2\nu_0})_{,r}+4r^{-1}e^{2\nu_0}\right] K^{(2)}_{,r} 
- 2(re^{2\nu_0})_{,r}H^{(2)}
+ 2 K^{(2)} \\
= \frac{-1}{2l +1} 
\Big\{
2r^2e^{2\nu_0}
\left[ \left( K^{(1)}_{,r} \right)^{2} -
H^{(1)}_{,r} K^{(1)}_{,r} \right]
+ \left( 2 \Lambda r^2 +l^2 +l -2 \right)
 \left( H^{(1)} \right)^{2} \\
- 2\left( l^2+l-2\right) H^{(1)} K^{(1)}
+ 2\left( l^2+l-1 \right) \left( K^{(1)} \right)^{2} 
\Big\},
\label{R2phiphi+R2thetatheta}
\end{multline}
where we have used the formulas
\begin{align}
\langle \left[P_l(\cos\theta)\right]^2\rangle&={1}/{(2l+1)},\\
\langle
\left[d{P}_l(\cos\theta)/d\theta\right]^2
\rangle &={l(l+1)}/{(2l+1)},
\end{align}
where
\begin{equation}
\langle f\rangle\equiv\frac12\int_0^\pi f(\theta)\sin\theta d\theta.
\label{theta_average}
\end{equation}
These equations come from $tt, rr$ components 
and the sum of the $\theta\theta, \phi\phi$ components 
of the Einstein equation, respectively.
The $r\theta$ component and 
the difference of the $\theta\theta, \phi\phi$ components 
provide no conditions for the $l=0$ mode. 
Similarly to the case of the first-order perturbation, these equations
do not overdetermine $H^{(2)}$ and $K^{(2)}$. 

The homogeneous solution 
for Eqs.~\eqref{R2tt}, \eqref{R2rr}, and \eqref{R2phiphi+R2thetatheta}
is given by
\begin{align}
&K^{(2)}_{hom}=C_1+C_2/r,\\
&H^{(2)}_{hom}=e^{-2\nu_0}
\left[C_1-{(e^{2\nu_0})_{,r}C_2}/{2}+{C_3}/{r}\right].
\end{align}
This is not physical because these perturbations 
are absorbed to the coordinate transformation
and changing the mass like
\begin{align}
&\bar{t}=(1-\epsilon^2C_1)t,\\
&\bar{r}=(1+\epsilon^2C_1)r+\epsilon^2C_2,\\
&\bar{m}=(1+3\epsilon^2C_1)m+\epsilon^2C_3. 
\end{align}
Moreover, $H^{(1)}_{hom}$ diverges on the horizon. 
Hence, in order to obtain a solution that does not diverge 
for $r_h\le r\le \infty$, we should impose the homogeneous
solution to be zero. This means that  
$K^{(2)}$ behaves like
\begin{equation}
K^{(2)}=\frac{d_2}{r^2}+\frac{d_3}{r^3}+\cdots,
\label{K2asymptotic}
\end{equation}
for large $r$, where $d_2, d_3,...$ are determined like
\begin{align}
&d_2=- \frac{a_{1}^{2}}{2 \left( 2l + 1 \right)} ,\\
&d_3= \frac{l(l+1) a_{1} c_{0}}{(2l+1)\Lambda } .
\label{asymptotic_d3}
\end{align}
The fact that $K^{(2)}=O(1/r^2)$ at large $r$ provides a
boundary condition for solving $K^{(2)}$ for $r_h<r<\infty$. 
The solution is written as
\begin{equation}
K^{(2)} = \frac{-1}{(2l+1)} 
\int_{r}^{\infty} \frac{1}{{r^{\prime\prime}}^{2}} 
\int_{r^{\prime\prime}}^{\infty}
\biggl[  {r^\prime}^{2} \left( K^{(1)}_{,r^\prime} \right)^{2}
- \frac{2l(l+1)}{e^{2\nu_0}} H^{(1)} \left(
H^{(1)} - K^{(1)} \right) \biggl] 
dr^{\prime} dr^{\prime\prime}.
\label{K2}
\end{equation}
We also see that $H^{(2)}$ behaves like
\begin{equation}
H^{(2)}=\frac{b_2}{r^2}+\frac{b_3}{r^3}+\cdots.
\label{H2asymptotic}
\end{equation}
Although $b_2$ is determined like
\begin{equation}
b_2=\frac{1}{2 \left( 2l + 1 \right)}
\left(
a_1^2-\frac{6c_0^2}{\Lambda}(l^2+l-1)
\right),
\end{equation}
it is not possible to determine $b_3$ in terms of $a_1$ and $c_0$,
because we cannot impose  $C_3$
of the homogeneous solution to be zero at infinity.
It is determined only by explicitly
solving $H^{(2)}$ with an appropriate boundary condition 
at the horizon $r=r_h$ which we introduce later.

Near the horizon,  the behavior of
$K^{(2)}$ is written as
\begin{equation}
K^{(2)}=\tilde{d}_0+\tilde{d_1}{(r-r_h)}+\cdots, 
\label{K2_horizon}
\end{equation}
where the coefficients are determined only by explicitly
calculating Eq.~\eqref{K2}:
\begin{align}
&\tilde{d}_0=
\frac{1}{(2l+1)}\left(
-\frac{c_0a_1}{r_h}
+\frac12\left(c_0^2-\tilde{c}_0^2\right)
+\int_{r_h}^{\infty}
\frac{(r-r_h)l(l+1)}{r_hre^{2\nu_0}}
H^{(1)}\left(2H^{(1)}-K^{(1)}\right)
dr
\right), \label{tilded0}
\\
& \tilde{d}_1=
\frac{1}{(2l+1)}
\left(
\frac{c_0a_1}{r_h^2}-2\tilde{c}_0\tilde{a}_1
-\int_{r_h}^{\infty}\frac{l(l+1)}{r_h^2e^{2\nu_0}}H^{(1)}
\left(2H^{(1)}-K^{(1)}\right)dr
\right).\label{tilded1}
\end{align}
Once $K^{(2)}$ is calculated, we find that
$H^{(2)}$ behaves like
\begin{equation}
H^{(2)}=\tilde{b}_0+\tilde{b_1}{(r-r_h)}+\cdots, 
\label{H2_horizon}
\end{equation}
near the horizon, where the coefficients are determined 
by Eqs.~\eqref{R2phiphi+R2thetatheta} and \eqref{R2rr} like
\begin{align}
&\tilde{b}_0=\frac{r_h}{2(3m-r_h)}
 \left(
(3m-r_h)\tilde{d}_1+\tilde{d}_0+\frac{l^2+l-1}{2l+1}\tilde{c}_0^2
\right),
\label{tildeb0}\\
&\tilde{b}_1=\frac12\tilde{d}_1+\frac{\Lambda r_h^2}{2(3m-r_h)}\tilde{b}_0.
\end{align}
The fact $H^{(2)}(r_h)=\tilde{b}_0$ gives a boundary condition for
solving $H^{(2)}$ for $r_h\le r\le\infty$. 
The solution of $H^{(2)}$ satisfying this boundary condition is written as
\begin{multline}
H^{(2)} = \frac{1}{2(2l+1)re^{2\nu_0}}
\int_{r_{h}}^{r} \bigg\{
(2l+1) 
\left[
2K^{(2)} 
+
\left({r^{\prime}}^2e^{2\nu_0}\right)_{,r^{\prime}}
   K^{(2)}_{,r^\prime}
\right] 
- {r^\prime}^2e^{2\nu_0}
K^{(1)}_{,r^\prime} 
\left(2H^{(1)}_{,r^\prime} -K^{(1)}_{,r^\prime}\right)
\\
+ \left( 2\Lambda {r^\prime}^{2} + 3l^2+3l-2 \right) 
\left( H^{(1)} \right)^{2}
- 2(l^2+l-1) K^{(1)}\left(2H^{(1)} -K^{(1)}\right) 
\bigg\} dr^\prime. 
\label{H2}
\end{multline}

Now we calculate the horizon area of the deformed black hole. 
The induced metric on the horizon $t={\rm const}$ and $r=r_h$
becomes
\begin{equation}
d\hat{s}^2=\left(
1+2\psi_1+2\psi_1^2+2\psi_2
\right)\big|_{r=r_h}
r_h^2\left(d\theta^2+\sin^2\theta d\phi^2\right). 
\end{equation}
Substituting Eqs.~\eqref{function_psi1} and \eqref{function_psi2}
and then using Eqs.~\eqref{K1_horizon} and \eqref{K2_horizon},
the change in the horizon area $\delta A$ is easily calculated:
\begin{eqnarray}
\frac{\delta A}{A} =
2 \epsilon^{2} \left( \frac{\tilde{c}_0^2}{2l + 1} 
+ \tilde{d}_0\right),
\label{A}
\end{eqnarray}
where $A\equiv\pi r_h^2$ is the area of 
the Schwarzschild$-$anti-de Sitter black hole. 
Because $\tilde{d}_0$ is given in Eq.~\eqref{tilded0},
we can calculate $\delta A$ without explicitly constructing solutions 
of $H^{(2)}$ and $K^{(2)}$. This quantity $\delta A/A$  provides
an indicator for the strength of nonlinear effect near the horizon.  

We also consider the area of the $t={\rm const}$ surface of the
conformal boundary $\mathcal{I}$ in the case where we choose the
conformal factor $\Omega=r^{-1}$,  
because it becomes the measure for the strength of nonlinear effect 
near infinity. We write this area and the increase in this area 
$S$ and $\delta S$, respectively. This is given by
 \begin{eqnarray}
\frac{\delta S}{S} =
2 \epsilon^{2} \left( \frac{c_0^2}{2l + 1} \right).
\label{S}
\end{eqnarray}
Because this is proportional to $c_0^2$ and is positive definite, this quantity
provides a natural factor for normalizing the thermodynamical quantities,
such as $\delta A/A$ and $\delta m/m$.

\section{Mass of the deformed black hole}

In this section, we calculate the mass of the deformed black hole. 
As we mentioned in the Introduction, we consider two definitions of mass:
the Abbott-Deser mass \cite{AD82} 
and the Ashtekar-Magnon mass \cite{AM84}. We will find that
these two definitions of mass give totally different results  
and will discuss the reason for this difference.

\subsection{Abbott-Deser mass}

Abbott and Deser found a conserved quantity for the spacetimes
with a cosmological constant~\cite{AD82}. 
They divided the metric tensor $\hat{g}_{\mu\nu}$
into two parts: 
\begin{equation}
\hat{g}_{\mu\nu}=\bar{g}_{\mu\nu}+h_{\mu\nu},
\end{equation}
where $\bar{g}_{\mu\nu}$ is the solution of the Einstein equation
which has the Killing vector field ${\bar{\xi}}_{\mu}$, and $h_{\mu\nu}$
represents a deviation from the background metric $\bar{g}_{\mu\nu}$. 
The conserved quantity associated with the Killing vector 
field $\bar{\xi}_{\mu}$ is
\begin{equation}
E(\bar{\xi})=\frac{1}{8\pi}\int
\sqrt{-\bar{g}} d^2x\left(
\bar{\xi}_{\nu}D_{\beta}K^{\mu\alpha\nu\beta}
-K^{\mu\beta\nu\alpha}D_{\beta}\bar{\xi}_{\nu}
\right)
D_{\mu}tD_{\alpha}r,
\end{equation}
where $\bar{g}$ is the determinant of $\bar{g}_{\mu\nu}$, 
$D_\mu$ denotes the
covariant derivative with respect to the background metric $\bar{g}_{\mu\nu}$,
$t$ and $r$ are the usual time and radial coordinates, and
$K^{\mu\alpha\nu\beta}$ is the superpotential defined by
\begin{equation}
K^{\mu\alpha\nu\beta}=
\frac12
\left(
\bar{g}^{\mu\beta}H^{\nu\alpha}+\bar{g}^{\nu\alpha}H^{\mu\beta}
-\bar{g}^{\mu\nu}H^{\alpha\beta}-\bar{g}^{\alpha\beta}H^{\mu\nu}
\right),
\end{equation} 
where 
\begin{equation}
H^{\mu\nu}=h^{\mu\nu}-\bar{g}^{\mu\nu}h^{\alpha}_{\alpha}/2.
\end{equation}
The indices are moving with respect to $\bar{g}_{\mu\nu}$, and
the integral is taken on the two-sphere at infinity $r=\infty$.
If $\bar{\xi}_{\mu}$ is the usual (past-directed) 
timelike Killing vector, $E(\bar{\xi})$ becomes
the total mass generated by $h_{\mu\nu}$ which is called the Abbott-Deser mass
$M_{AD}$. 

Now we calculate the Abbott-Deser mass of the deformed black hole.
We adopt the Schwarzschild$-$anti-de Sitter spacetime given by 
Eqs.~\eqref{SAD1} and \eqref{SAD2} as the background spacetime. 
By a  straightforward calculation, we find that
the Abbott-Deser mass of the spacetime of Eq.~\eqref{metric_deformed}
becomes
\begin{equation}
M_{AD}=\lim_{r\to\infty}\frac{r^2}{8\pi}\int 
\left[ 
e^{2\nu_0}
\left(
\frac1r\left(e^{2(\mu-\mu_0)}-e^{2\psi}\right)
-(e^{2\psi})_{,r}
\right)
+\frac12(e^{2\nu_0})_{,r}(e^{2\psi}-1)
\right]d\Omega, 
\label{AD_mass}
\end{equation} 
where $d\Omega\equiv \sin\theta d\theta d\phi$ and
we used $\bar{\xi}^{\mu}=(-1, 0, 0, 0)$.
By expanding this formula in $\epsilon$
using Eqs.~\eqref{expand_mu} and \eqref{expand_psi}, 
$M_{AD}$ is written like
\begin{equation}
M_{AD}=\epsilon M^{(1)}_{AD}+\epsilon^2 M^{(2)}_{AD}+\cdots. 
\end{equation}
Because $\mu_1$ and $\psi_1$ are proportional to $P_l(\cos\theta)$,
the integration in Eq.~\eqref{AD_mass} immediately 
leads to $M^{(1)}_{AD}=0$ for $l\ge 2$. 
For the second-order Abbott-Deser mass $M^{(2)}_{AD}$, 
we find, by substituting 
Eqs.~\eqref{function_nu1_mu1}, \eqref{function_psi1}, \eqref{function_mu2},
and \eqref{function_psi2}
and then using 
Eqs.~\eqref{H1asymptotic}, \eqref{K1asymptotic}, \eqref{K2asymptotic},
and \eqref{H2asymptotic}, 
\begin{equation}
M^{(2)}_{AD}=\lim_{r\to\infty}
\left(\frac{2\Lambda c_0a_1}{3(2l+1)}r^2+\cdots\right)=-\infty.
\label{AD_result}
\end{equation}
Thus the Abbott-Deser mass diverges to minus infinity at second order.

One interpretation for this divergence of the Abbott-Deser mass is that
the deformed black hole is a far-lower-energy state compared
to the Schwarzschild$-$anti-de Sitter black hole and 
the Schwarzschild$-$anti-de Sitter black hole would rapidly
deform toward the lower-energy state.  
However, we would like to point out that such an interpretation
may not be correct. To see this, 
we consider the coordinate transformation $r\to r+\epsilon a$ of the
Schwarzschild$-$anti-de Sitter metric.
The resulting metric is given by Eq.~\eqref{metric_deformed} with
\begin{align}
&e^{-2\nu}=e^{2\mu}=e^{2\mu_0}
+\left(e^{2\mu_0}\right)_{,r}a\epsilon
+\left(e^{2\mu_0}\right)_{,rr}a^2\epsilon^2/2+\cdots,\\
&e^{2\psi}=1+{2a}\epsilon/r+{a^2}\epsilon^2/r^2.
\end{align}
Substituting these formulas to Eq.~\eqref{AD_mass}, we find $M^{(1)}_{AD}=0$
and
\begin{equation}
M^{(2)}_{AD}=\lim_{r\to\infty}\left(-\frac{5\Lambda a^2}{6}r+\cdots\right)
=\infty.
\label{AD_coordinate}
\end{equation}
Thus the Abbott-Deser mass for the perturbation generated by the 
coordinate transformation also diverges at second order. 
This indicates that the Abbott-Deser mass
is not gauge invariant at second order and hence
its divergence may be spurious. In fact,
we can obtain a finite mass for
the deformed black hole,  if we 
use the covariant definition of mass given by Ashtekar and Magnon.

\subsection{Ashtekar-Magnon mass}

Ashtekar and Magnon constructed a conserved quantity of a
weakly asymptotically anti$-$de Sitter spacetime 
$(\hat{M}, \hat{g}_{\mu\nu})$~\cite{AM84}. 
They considered the conformally transformed spacetime $(M, g_{\mu\nu})$
where $g_{\mu\nu}=\Omega^2\hat{g}_{\mu\nu}$ and the situation that
$M$ has a boundary 
$\mathcal{I}$ whose topology is $R\times S^2$. 
Their conserved quantity is defined on the spacelike $S^2$ surface 
(denoted by $\mathcal{C}$) on the 
conformal boundary $\mathcal{I}$ as follows:
\begin{equation}
Q_{\xi}\equiv -\frac{1}{8\pi}\left(-\frac{3}{\Lambda}\right)^{3/2}
\int_{\mathcal{C}}\Omega^{-1}C_{\alpha\mu\beta\nu}
\xi^{\alpha} n^{\mu}N^{\beta}n^{\nu}dS,
\end{equation}
where $n_{\mu}=\nabla_{\mu}\Omega$, $N^{\beta}$ is the timelike unit normal
on $\mathcal{C}$, $\xi^{\alpha}$ is the conformal Killing vector field 
on $\mathcal{I}$, 
$dS$ is the proper area element of $\mathcal{C}$, and
$C_{\alpha\mu\beta\nu}$ denotes the Weyl tensor of $M$. All index moving
and covariant derivatives are with respect to the metric of the conformally
transformed spacetime $(M, g_{\mu\nu})$. 
This definition is conformally invariant with the same choice of
the coordinate components of $\xi^{\mu}$. 
If $\xi^\mu$ is the (future-directed) timelike Killing vector field, 
this conserved quantity
is the Ashtekar-Magnon mass $M_{AM}$ of the spacetime. 
For the axisymmetric spacetime with metric~\eqref{metric_deformed},
we find by a straightforward calculation that
the Ashtekar-Magnon mass becomes
\begin{multline}
M_{AM}=\lim_{r\to\infty}-\sqrt{\frac{-3}{16\Lambda}}
\int_{0}^{\pi}
\xi^t r^2e^{\nu}
\left(
e^{-2\mu}\left[\nu_{,rr}+\nu_{,r}(\nu_{,r}-\mu_{,r})\right]
+\frac{e^{-2\psi}}{r^2}\mu_{,\theta}\nu_{,\theta}+\frac{\Lambda}{3}
\right)
e^{2\psi}\sin\theta d\theta.
\label{AM_mass}
\end{multline} 
If we calculate this quantity for the Schwarzschild$-$anti-de Sitter spacetime,
we obtain $M_{AM}=m$ choosing $\xi^t=1$. We can easily show that
this quantity does not change 
under the coordinate transformation $r\to r+\epsilon a$.

In the above definition,  how to choose the norm of $\xi^{\mu}$
has not been specified. Here we would like to discuss this criterion
using one concrete example, 
because the mass calculation of the deformed black hole crucially
depends on the choice of the norm of $\xi^{\mu}$. 
If we make a coordinate transformation $r\to ar$ and $t\to bt$
to the Schwarzschild$-$anti-de Sitter spacetime, the resulting metric 
becomes Eq.~\eqref{metric_deformed} with
\begin{align}
b^{-2}e^{2\nu}=a^2e^{-2\mu}&=1-2m/ar-\Lambda a^2r^2/3,\\
e^{2\psi}&=a^2.
\end{align}
Choosing $\Omega=1/r$, the metric of the
conformal boundary $\mathcal{I}$ becomes
\begin{equation}
ds^2=\left(\Lambda a^2b^2/3\right)dt^2
+a^2\left(d\theta^2+\sin^2\theta d\phi^2\right).
\end{equation}
Calculating Eq.~\eqref{AM_mass} for this metric, we obtain $M_{AM}=\xi^t bm$.
Because the mass should be invariant under the coordinate transformation,
this means that we should choose $\xi^t=b^{-1}$. One of the natural
general criteria that recover this choice is as follows: 
if there is a conformal transformation such that the norm of the
Killing vector field $\xi^{\mu}$ becomes constant on $\mathcal{I}$, 
we should choose $\xi^{\mu}$ that satisfy
\begin{equation}
\left({3/}{\Lambda}\right)\xi^\mu\xi_\mu={S}/{4\pi},
\label{choosing_norm}
\end{equation}
for the calculation of the Ashtekar-Magnon mass, 
where $S$ denotes the proper area of 
the $t={\rm const}$ surface on $\mathcal{I}$. 
We adopt this criterion in calculating the Ashtekar-Magnon mass
of the deformed black hole.

Now we calculate the Ashtekar-Magnon mass of the deformed black hole.
From the above discussion, we use $\xi^t$ 
in the calculation of Eq.~\eqref{AM_mass} as follows:
\begin{equation}
\xi^t=\lim_{r\to\infty}\sqrt{\langle e^{2\psi}\rangle}
=\lim_{r\to\infty}1+
\epsilon^2\left(
\langle\psi_2\rangle+\langle\psi_1^2\rangle
\right)+O(\epsilon^4),
\end{equation}
where the definition of $\langle f\rangle$
is given in Eq.~\eqref{theta_average}.
Substituting this formula and
expanding Eq.~\eqref{AM_mass} using 
Eqs.~\eqref{expand_nu}, \eqref{expand_mu}, and \eqref{expand_psi}, 
$M_{AM}$ can be written like
\begin{equation}
M_{AM}=M^{(0)}_{AM}+\epsilon M^{(1)}_{AM}+\epsilon^2 M^{(2)}_{AM}+\cdots. 
\end{equation}
We immediately obtain $M^{(0)}_{AM}=m$ and $M^{(1)}_{AM}=0$ for $l\ge 2$. 
Substituting 
Eqs.~\eqref{function_nu1_mu1}, \eqref{function_psi1}, \eqref{function_nu2},
\eqref{function_mu2}, and \eqref{function_psi2}
and then using 
Eqs.~\eqref{H1asymptotic}, \eqref{K1asymptotic}, \eqref{K2asymptotic},
and \eqref{H2asymptotic} for $M^{(2)}_{AM}$, 
we obtain, after a rather lengthy calculation,
\begin{equation}
M^{(2)}_{AM}
=\frac13\Lambda\left(\frac{2\left(a_1a_2-a_3c_0\right)}{2l+1}-b_3\right)
+\frac{2a_1c_0}{2l+1}
+\frac{3mc_0^2}{2l+1}.
\label{AM_result_1}
\end{equation} 
This is finite in contrast to the Abbott-Deser mass. 
Because the Ashtekar-Magnon mass is a covariant definition of mass
which is welldefined even in the weakly asymptotically anti$-$de Sitter
spacetimes, we can regard it as a real amount of energy contained 
in the spactimes with deformed black holes. 
The validity of the Ashtekar-Magnon mass is also supported
by the expectation that the continuous change of mass should result
from the continuous deformed black hole series. 
Hence we consider that there is a possibility 
of the quasistatic deformation of the spacetime boundaries in weakly
asymptotically anti$-$de Sitter spacetimes and  
we investigate the thermodynamic law of the deformed black holes 
with this mass definition in the following two sections.

To simplify Eq.~\eqref{AM_result_1}, we consider the method of
expressing $b_3$ with $H^{(1)}$ and $K^{(1)}$.
By summing Eqs.~\eqref{R2tt} and \eqref{R2rr} and then 
rewriting the right-hand side using Eqs.~\eqref{Rtt} and \eqref{Rr theta}, 
we find
\begin{multline}
\left(r^2e^{2\nu_0}H^{(2)}_{,r}\right)_{,r}
+2mH^{(2)}_{,r}
-\frac23\Lambda\left(r^3H^{(2)}\right)_{,r}
-\left(r^2e^{2\nu_0}K^{(2)}_{,r}\right)_{,r}\\
=\frac{1}{2l+1}
\left[
\left(r^2e^{2\nu_0}H^{(1)}H^{(1)}_{,r}\right)_{,r}
-l(l+1)\left(H^{(1)}\right)^2
\right].
\end{multline}
Integrating this equation from the horizon to large $r$ and
substituting Eqs.~\eqref{H1asymptotic}, \eqref{K2asymptotic}, 
and \eqref{H2asymptotic},
we obtain
\begin{equation}
\frac{\Lambda}{3}b_3=\frac{5l(l+1)-6}{2(2l+1)}a_1c_0
+2(3m-r_h)\tilde{b}_0
-\frac{l(l+1)}{2l+1}\int_{r_h}^{\infty}\left(H^{(1)}\right)^2dr.
\end{equation}
Substituting this formula, Eqs.~\eqref{a3} and \eqref{c0} 
into Eq.~\eqref{AM_result_1}, we finally find
\begin{equation}
M^{(2)}_{AM}=
\frac{-(l^2+l+2)}{2(2l+1)}a_1c_0
-2(3m-r_h)\tilde{b}_0
+\frac{l(l+1)}{2l+1}\int_{r_h}^{\infty}\left(H^{(1)}\right)^2dr
+\frac{3mc_0^2}{2l+1},
\label{AM_result_2}
\end{equation}
where $\tilde{b}_0$  
can be written in terms of $\tilde{d}_0$ and $\tilde{d}_1$
using Eq.~\eqref{tildeb0}, 
which in turn are
given in Eqs.~\eqref{tilded0} and \eqref{tilded1}.
Hence we can express $M^{(2)}_{AM}$ in terms of the quantities
of only the first-order perturbation.

\section{Approximate calculation for $\alpha\gg 1$ and $\alpha\ll 1$}

As we mentioned in the Secs. I and II, to solve analytically
the equations of the perturbation is rather difficult. 
However, we can construct the solution approximately
in the two simple cases: one is the case that the
anti$-$de Sitter radius is much larger than the Schwarzschild radius 
$\alpha \ll 1$, and the other is that the anti$-$de Sitter radius
is much smaller than the Schwarzschild radius $\alpha \gg 1$. 
In this section, we will show the solution for these two cases
and calculate the horizon area and the Ashtekar-Magnon mass.
Using these results, we can discuss the properties
of the first law of the black hole thermodynamics for the
deformed black holes.

\subsection{$\alpha \ll 1$ case}

Setting a new coordinate $x\equiv r/2m$, the function $e^{2\nu_0}$ becomes
\begin{equation}
e^{2\nu_0}= 1 - {1}/{x} + \left( \alpha x \right)^{2}. 
\end{equation}
In the case of $\alpha\ll 1$, the horizon location is $x=x_h\simeq 1$.
In the region $1\le x\lesssim \alpha^{-2/3}$, the order of the second term 
is larger than $O(\alpha^{2/3})$ and the order of the third term
is smaller than  $O(\alpha^{2/3})$. 
Hence the sum of the first and second terms
is much larger than the third term: the spacetime is Schwarzshild-like. 
In the region $\alpha^{-2/3}\lesssim x\le \infty$, the order of the second 
term is smaller than $O(\alpha^{2/3})$ and the order of the third term
is greater than $O(\alpha^{2/3})$. Because the sum of the first and 
third terms is much greater than the second term, the spacetime 
is similar to the anti$-$de Sitter spacetime.
Hence, we can use the matching method: we construct solutions for
the Schwarzschild spacetime and the anti$-$de Sitter spacetime
in the regions $1\le x\lesssim \alpha^{-2/3}$ and 
$\alpha^{-2/3}\lesssim x\le \infty$,
respectively, and then match these two solutions in the
overlapping region $x \sim \alpha^{-2/3}$.

The general solutions of $H^{(1)}$ for Schwarzschild spacetime
are given in \cite{RW57}. 
The solutions that satisfy the boundary condition on the horizon
can be expressed in terms of associated Legendre's polynomials,
$H^{(1)}_{in}=P_l^2(2x-1)$. 
To construct a solution $H^{(1)}$ in the anti$-$de Sitter regime,
we introduce a new coordinate $y\equiv \sqrt{-\Lambda/3}r$.
Setting $Y\equiv y^2$ and $M=Y^{(l+2)/2}\tilde{M}$, Eq.~\eqref{H1} in the
anti$-$de Sitter regime becomes
\begin{equation}
Y(1+Y)\tilde{M}_{,YY}+
\left[
(l+1/2)Y+(l+3/2)
\right]\tilde{M}_{,Y}
+\left[l(l-1)/4\right]\tilde{M}=0.
\end{equation}
Because this equation is related to the hypergeometric equation,
one of the solutions becomes
$\tilde{M} = 
{}_2F_1 \left( l/2 ,\ (l-1)/2; \ l + {3}/{2} ; \ -Y \right),$
and the corresponding solution is
\begin{equation}
H^{(1)}_{out}=\frac{y^l}{y^2+1}
{}_2F_1 \left( l/2 ,\ (l-1)/2; \ l + {3}/{2} ; \ -y^2 \right),
\label{H1_out}
\end{equation}
which behaves like $H^{(1)}_{out}\simeq y^l$ in the matching region. 
Since the behavior of $H^{(1)}_{in}$ is proportional to 
$x^l$ in the overlapping region,
we can match these two solutions.  The solution of $H^{(1)}_{in}$
that smoothly continues to Eq.~\eqref{H1_out} can be written as
\begin{equation}
H^{(1)}_{in}=-
\left(\frac{\alpha}{2}\right)^l
\frac{(l-2)!}{(2l-1)!!}P_l^2(2x-1).
\label{H1_in}
\end{equation}
Although the convergence region of Eq.~\eqref{H1_out} is $0\le y\le 1$,
we can make an analytic continuation to the region $1\le y\le \infty$
using well-known techniques. Hence we have constructed the solution of
$H^{(1)}$ in the $\alpha\ll 1$ case. Using Eq.~\eqref{K1},
we can also write down the solutions $K^{(1)}_{in}$ 
and $K^{(1)}_{out}$ in terms of associated Legendre's polynomials and
the hypergeometric functions, respectively, 
although we do not show them explicitly here. Using this solution of 
$K^{(1)}$ or observing the behavior of $H^{(1)}_{in}$ and $H^{(1)}_{out}$
near the horizon and infinity and then using Eqs.~\eqref{c0} and
\eqref{tilde_c0},   we find
\begin{equation}
\frac{\tilde{c}_0}{c_0}=
\frac{\Gamma((l-1)/2)\Gamma((l+3)/2)\Gamma(l+3)}{4\Gamma(l+3/2)\Gamma(l+1/2)}
\left(\frac{\alpha}{4}\right)^l.
\end{equation}
Hence, for $\alpha\ll 1$, 
the deformation of the horizon is much smaller than that of
spacelike infinity, and this tendency is enhanced  
for larger $l$.

Now that we have constructed a solution of the first-order perturbation,
we would like to calculate the mass and horizon area. However, 
this calculation for general $l$
requires integration of the products of the 
hypergeometric functions, and the analytic calculation is rather difficult.
Hence we consider analytically only the $l=2$ case, 
where the function in Eq.~\eqref{H1_out} reduces to
the elementary functions. For $l>2$, we numerically evaluate the
relation of the area and mass.
Before doing this, we would like to remark on the
general properties of the second-order solution in the $\alpha\ll1$ case. 
As we can see from Eqs.~\eqref{K2} and \eqref{H2}, 
the order of the second-order perturbation 
becomes $O(\alpha^{-1}\epsilon^2)$, which is
much larger than $\epsilon^2$. 
This indicates that the nonlinear effect 
rapidly increases with an increase in $\epsilon$.
This perturbative analysis is only reliable for a sufficiently 
small $\epsilon$ that satisfies $\epsilon < \alpha$.

Now we calculate the horizon area and the mass for the $l=2$ case.
The solutions of $H^{(1)}_{out}$ and $K^{(1)}_{out}$ corresponding
to Eq.~\eqref{H1_out} can be rewritten as
\begin{align}
& H^{(1)}_{out}=\frac{5}{8y^3\left(y^2+1\right)}
\left[
-y\left(5y^2+3\right)+3\left(y^2+1\right)^2\arctan y
\right],\\
& K^{(1)}_{out}=\frac{5}{2}-\frac{15}{8y^3}
\left[
y+\left(y^2-1\right)\arctan y
\right].
\end{align}
Using Eqs.~\eqref{A}, \eqref{S}, and \eqref{AM_result_2}, we find
\begin{align}
&\frac{\delta A}{A} \simeq 
\frac{3\pi}{8}
\left(12\log2 -11\right)\alpha^{-1}\epsilon^2,\\
&\frac{\delta m}{m}\simeq 
\frac{3\pi}{4}
\left(3\log2 -4\right)\alpha^{-1}\epsilon^2,\\
&\frac{\delta S}{S}\simeq 
\frac{5}{2}\epsilon^2,
\end{align}
where $\delta m\equiv M^{(2)}_{AM}\epsilon^2$. 
Note that both $\delta A$ and $\delta m$ are negative. 
Because $(\delta A/A)/(\delta S/S)=O(\alpha^{-1})$,
the back reaction to the horizon area 
is far larger than that to the spacelike area at infinity. 
This result implies 
\begin{equation}
\frac{(\delta m/\delta A)}{(\kappa/8\pi)}=
\frac{12\log 2-16}{12\log 2-11}\simeq 2.86,
\label{thermodynamic_alpha_ll_1_coeffecient}
\end{equation}
which indicates that the first law in the usual form does not hold
for deformed black holes. In other words, the first law for
the deformation of the Schwarzschild$-$anti-de Sitter black holes 
has a correction term such that 
\begin{eqnarray}
\delta m = \frac{\kappa}{8 \pi} \delta A + \delta W,
\label{thermodynamic_alpha_ll_1}
\end{eqnarray}
where $\delta W$ can be expressed in terms of the coefficients of
$H^{(1)}$ and $K^{(1)}$ as $\delta W=-a_1c_0\epsilon^2/5$
in the $\alpha\ll 1$ case.
Because $c_0$ is related to the deformation of spacelike infinity,
this term would be related to the work which is necessary for the
deformation of infinity. In the next section, we will discuss
the origin of this work term in more detail.

Because the work term $\delta W$ in Eq.~\eqref{thermodynamic_alpha_ll_1} 
is negative, the change in the horizon area $\delta A$
of the deformed black hole 
and that of the Schwarzschild$-$anti-de 
Sitter black hole $\delta {A_0}$ for the same $\delta m$
(i.e., $\kappa\delta A_0/8\pi=\delta m$) 
satisfy the relation $\delta A_0<\delta A$.
This implies that the area of the
deformed black hole is larger than that of the Schwarzschild$-$anti-de Sitter
black hole if compared under the same Ashtekar-Magnon mass. 
Interestingly, we can claim that 
the deformation of the Schwarzschild$-$anti-de Sitter black hole 
is a process consistent with the area theorem. 
Although the area theorem has not been proved in the spacetime
with $\Lambda<0$ and there are some counterexamples for this theorem
such as the black holes in Brans-Dicke theory \cite{Kan96},
we expect that this result indicates the importance of the
solution sequence of the deformed black hole.

\begin{table}[tb]
\caption{ The value of $\alpha (\delta A/A)/(\delta S/S)$, 
$\alpha (\delta m/m)/(\delta S/S)$, and 
$\alpha (\delta W/m)/(\delta S/S)$ for $l=2,...,9$.
The work term is negative for all $l$ and  
it becomes large as $l$ increases. 
}
\begin{ruledtabular}
\begin{tabular}{c|cccccccc}
  $l$ & 2 & 3 & 4 & 5 & 6 & 7 & 8 & 9 \\
  \hline 
  $\alpha (\delta A/A)/(\delta S/S)$ 
& $-1.26$ & $-1.86$ & $-2.45$ & $-3.02$ & $-3.60$ & $-4.17$ & $-4.73$ & $-5.30$\\
  $\alpha (\delta m/m)/(\delta S/S)$ 
& $-1.81$ & $-5.18$ & $-11.2$ & $-20.5$ & $-34.0$ & $-52.4$ & $-76.4$ & $-107.$\\
  $\alpha (\delta W/m)/(\delta S/S)$ 
& $-1.18$ & $-4.24$ & $-9.94$ & $-19.0$ & $-32.2$ & $-50.3$ & $-74.0$ & $-104.$\\
\end{tabular}
\end{ruledtabular}
\end{table}

Now we discuss the $l>2$ cases. Calculating the integrals in 
Eqs.~\eqref{A} and \eqref{AM_result_2} numerically,
we evaluate the factor $\alpha(\delta A/A)/(\delta S/S)$,
$\alpha(\delta m/m)/(\delta S/S)$, and $\alpha(\delta W/m)/(\delta S/S)$. 
In all cases, $\delta A$, $\delta m$, and
$\delta W$ are negative. 
The results are shown in Table I. 
The work term estimated by $(\delta W/m)/(\delta S/S)$ 
increases for larger $l$
and the difference from the ordinary thermodynamical 
relation becomes large. 
A higher-multipole deformation requires a larger work term. 
Interestingly, we have found that the relation
\begin{equation}
\delta W=-\frac{(l-1)(l+2)}{4(2l+1)}a_1c_0\epsilon^2
\label{work_a1c0}
\end{equation}
holds with accuracy $10^{-6}$. This indicates that this would be
the exact value and that our numerical calculation is accurate. 
Similarly to the $l=2$ case, the area of the deformed
black hole is larger than that of the Schwarzschild$-$anti-de Sitter
black hole if compared under the same mass.

\subsection{$\alpha \gg 1$ case}

In the case of $\alpha\gg 1$, we have the relation 
$x_h\simeq \alpha^{-2/3}\ll 1$ for the location of the black hole 
$x=x_h$.
Near the horizon, 
the first term of $e^{2\nu_0}=1-1/x+(\alpha x)^2$  
is $\alpha^{-2/3}$ times smaller compared to the other two terms.
In the region $x\gtrsim 1$,  the third term $(\alpha x)^2$ in $e^{2\nu_0}$
is more than $\alpha^2$ times larger than the other two terms. 
Hence, for all $x_h\le x\le\infty$, the first term
is small compared to the sum of the other two terms
and we neglect it in the following analysis.
In this approximation, we should neglect terms whose order
is $\alpha^{-2/3}$ times the leading order. 

Setting $z\equiv r/r_h$ and $Z\equiv z^3$,  
Eq.~\eqref{H1} is written as
\begin{eqnarray}
3Z \left( Z - 1 \right) {M}_{,ZZ} 
-\left( 2Z+1 \right) {M}_{,Z} +
{2} M = 0,
\end{eqnarray}
under the above approximation. 
The solution satisfying the boundary condition becomes
$M=2Z-3Z^{2/3}+1$, and the corresponding solution $H^{(1)}$
becomes 
\begin{eqnarray}
H^{(1)} 
= - \frac{1}{z} + \frac{3z}{z^{2} + z + 1}+O(\alpha^{-2/3}).
\end{eqnarray}
Using Eq.~\eqref{K1}, we derive
\begin{eqnarray}
K^{(1)} &=& \frac{6 \alpha^{2/3}}{l^2+l-2} + O(1).
\end{eqnarray}
In summary, the solutions of $H^{(1)}$ and $K^{(1)}$ are
\begin{align}
&H^{(1)}=O(\alpha^{-2/3}),
\label{H1_alpha_gg_1}\\
&K^{(1)}=\left(2l+1\right)^{1/2}+O(\alpha^{-2/3}),
\label{K1_alpha_gg_1}
\end{align}
where we used the degree of freedom to choose the amplitude of $K^{(1)}$.
It is a remarkable fact that $K^{(1)}$ is almost constant and hence 
$|\tilde{c}_0|\simeq |c_0|$.
Although we mentioned in Sec. II that the conformal boundary 
at infinity is more deformed compared to the horizon, 
the deformations of the two are similar if 
$|\Lambda|$ is large. 
This feature does not depend on $l$. 
These results are in contrast to the  $\alpha\ll 1$ case.

Substituting Eqs.~\eqref{H1_alpha_gg_1} and \eqref{K1_alpha_gg_1} into 
Eqs.~\eqref{A}, \eqref{S}, and \eqref{AM_result_2}, we immediately find
\begin{align}
\delta A/A&=\left[{2}+O(\alpha^{-2/3})\right]\epsilon^2,\\
\delta S/S&=\left[{2}+O(\alpha^{-2/3})\right]\epsilon^2,\\
\delta m/m&=\left[{3}+O(\alpha^{-2/3})\right]\epsilon^2. 
\end{align}
In this case, the back reaction to the horizon area is similar to
that to the spacelike area at infinity. 
This result leads to 
\begin{eqnarray}
\delta m \simeq \left({\kappa}/{8 \pi}\right) \delta A,
\label{thermodynamic_alpha_gg_1}
\end{eqnarray}
which means that the thermodynamic law of 
the deformed black holes is almost the same as that of the
Schwarzschild$-$anti-de Sitter black holes if $|\Lambda|$ is sufficiently large.
In contrast to the $\alpha\ll 1$ case, the work term is small 
in the $\alpha\gg 1$ case.  
Hence we find that there is a correlation between the value of the
work term and the difference of the 
deformation of the horizon and the spacelike surface at infinity: 
the absolute value of the work term decreases 
as $|\tilde{c}_0/c_0|$ approaches unity.

\section{Numerical calculation for $\alpha\sim 1$}

In this section, we numerically investigate the deformed black hole 
for $\alpha\sim 1$ to complete the perturbative analysis. 
Equations~\eqref{Rtt}, \eqref{Rrr}, \eqref{Rr theta}, and \eqref{R phi phi}
can be reduced to two first-order differential equations
for $(rH^{(1)})$ and $K^{(1)}$, which asymptote to constant 
values for large $r$. We solved these equations using
the Runge-Kutta method from the horizon $r=r_h$ to the cutoff value 
$r=r_c$. We selected the grid number and the cutoff value $r=r_c$
as follows.  Because there are two characteristic length scales 
 $R_A$ and $R_S$, 
we set $10^2$ grids within the smaller length scale 
and solved in a range which is $10^2$ times as long as 
the larger length scale. 
Hence, the cutoff value is $r_c=r_h+10^2\times{\rm max}[R_A, R_S]$
and the total grid number becomes $10^4\times{\rm max}[\alpha, \alpha^{-1}]$.
Beyond the cutoff $r=r_c$, 
we approximate $H^{(1)}$ and $K^{(1)}$ with the formulas     
$H^{(1)}\simeq a_1/r+a_2/r^2$ and $K^{(1)}\simeq c_0+c_1/r$, 
from which we determine the value of $c_0$ and $a_1$ using 
Eq.~\eqref{c0}, Eq.~\eqref{c1}, 
and the numerical values of $(rH^{(1)})$ and $K^{(1)}$ 
at $r=r_c$, and evaluate the integrals in Eqs.~\eqref{A} and
\eqref{AM_result_2} beyond the cutoff $r>r_c$.
The numerical error is about $0.1$\%, 
which is estimated by using several different grid numbers 
and the cutoff values. 
In the $\alpha=10^{-3}$ case, our numerical results coincide with 
the values in Table I with $0.1$\% accuracy.

\begin{figure}[tb]
\centering
{
\includegraphics[width=0.6\linewidth]{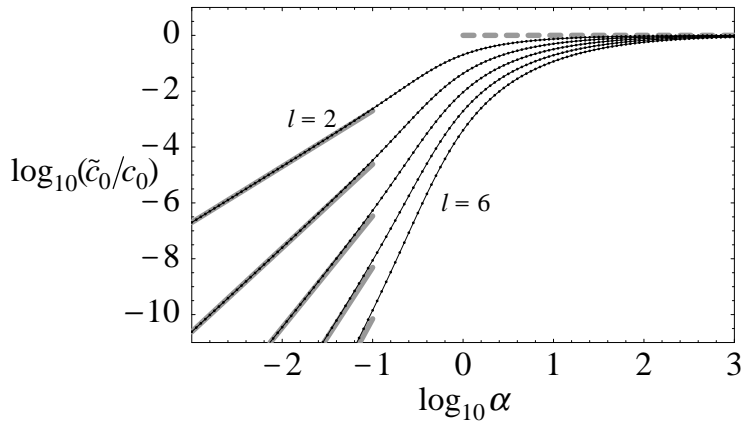}
}
\caption{The relation of $\alpha$ and the ratio of 
the deformation of the horizon and spacelike infinity $\tilde{c}_0/c_0$
for $l=2,...,6$. Both axes are shown in the $\log$ scale.
The gray dashed line shows the asymptotic behavior for $\alpha\gg 1$ and
the gray solid line shows the asymptotic 
behavior for $\alpha\ll 1$ for each $l$. }
\end{figure}

\begin{figure}[tb]
\centering
{
\includegraphics[width=0.6\linewidth]{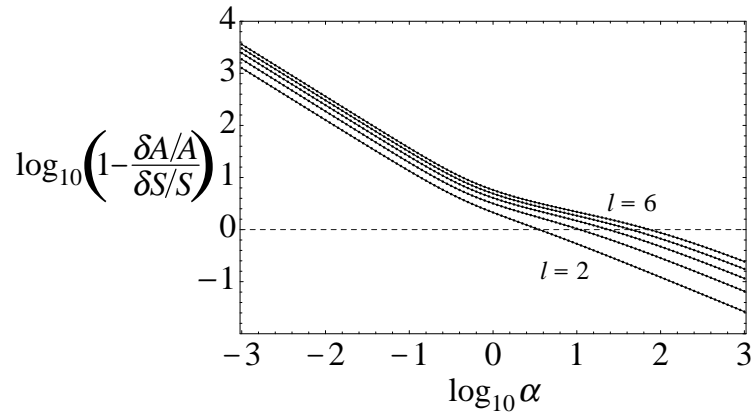}
}
\caption{The relation of $\alpha$ and change in the horizon area 
$1-(\delta A/A)/(\delta S/S)$ for $l=2,...,6$. 
Both axes are shown in the $\log$ scale.
The value of $1-(\delta A/A)/(\delta S/S)$ 
is almost proportional to $\alpha^{-1}$ for $\alpha\ll 1$ and
is proportional to $\alpha^{-2/3}$ for $\alpha\gg 1$. 
The location where $\delta A$ becomes zero is shown by a dashed line.}
\end{figure}

\begin{figure}[tb]
\centering
{
\includegraphics[width=0.6\linewidth]{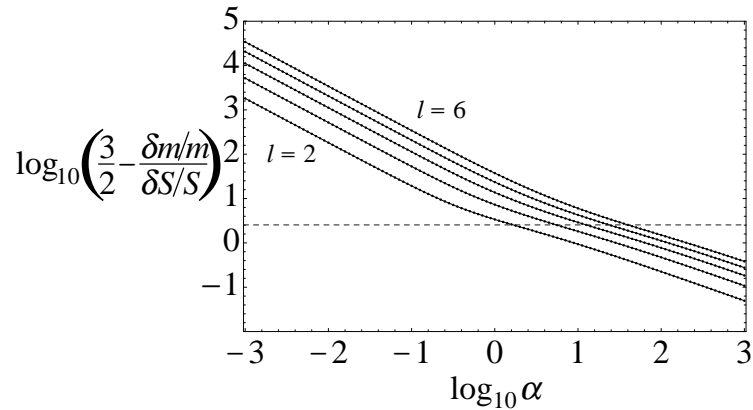}
}
\caption{The relation of $\alpha$ and the change in the mass
$3/2-(\delta m/m)/(\delta S/S)$ for $l=2,...,6$.  
Both axes are shown in the $\log$ scale.
The value of $3/2-(\delta m/m)/(\delta S/S)$ 
is almost proportional to $\alpha^{-1}$ for $\alpha\ll 1$ and
is proportional to $\alpha^{-2/3}$ for $\alpha\gg 1$. 
The location where $\delta m$ becomes zero is shown by a dashed line.}
\end{figure}

Now we show the numerical results. Figure 1 shows the behavior of 
the ratio $(\tilde{c}_0/c_0)$ of 
the deformation of the two spacetime boundaries---i.e.,  
the horizon and the two-surface at spacelike infinity---as 
a function of $\alpha$. The asymptotic behaviors for $\alpha\gg 1$ and
$\alpha\ll 1$ derived in Sec. IV are also shown. 
We see that the matching method
gives a fairy good approximation for $\alpha\lesssim 10^{-1}$. 
The behavior of $1-(\delta A/A)/(\delta S/S)$
and $3/2-(\delta m/m)/(\delta S/S)$ are shown in Figs. 2 and 3, 
respectively.   
We see that both $(\delta A/A)/(\delta S/S)$ and $(\delta m/m)/(\delta S/S)$ 
are proportional to $\alpha^{-1}$ for
$\alpha\ll 1$, which is consistent with the analysis of Sec. IV. 
In the region $\alpha\gg 1$, 
$1-(\delta A/A)/(\delta S/S)$ and $3/2-(\delta m/m)/(\delta S/S)$ asymptote
to zero and these values are proportional to $\alpha^{-2/3}$. This is also
consistent with the results in Sec. IV.  

\begin{figure}[tb]
\centering
{
\includegraphics[width=0.6\linewidth]{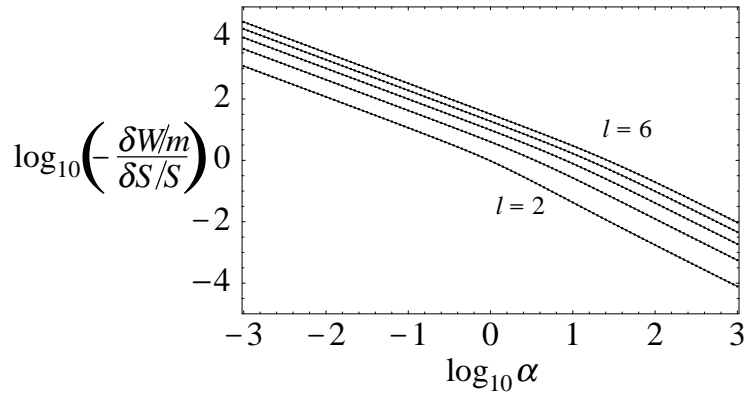}
}
\caption{The relation of $\alpha$ and the work term 
$-(\delta W/m)/(\delta S/S)$ for $l=2,...,6$.  
Both axis are shown in the $\log$ scale.  
The value of $-(\delta W/m)/(\delta S/S)$ 
is almost proportional to $\alpha^{-1}$ for $\alpha\ll 1$ and
is proportional to $\alpha^{-4/3}$ for $\alpha\gg 1$. }
\end{figure}

Figure 4 shows the behavior of the work term $(\delta W/m)/(\delta S/S)$. 
We see that the value of $\log_{10}[-(\delta W/m)/(\delta S/S)]$
does not diverge and hence $\delta W$ always takes a negative value.
This implies that the area of the deformed black hole is larger than that
of the Schwarzschild$-$anti-de Sitter black hole if compared under the same 
Ashtekar-Magnon mass for arbitrary $\alpha$. 
In the region $\alpha\ll 1$, the value of 
$(\delta W/m)/(\delta S/S)$ is proportional to $\alpha^{-1}$, which coincides
with the results of Sec. IV. 
This behavior of $(\delta W/m)/(\delta S/S)$ indicates 
that quasistatic deformation requires a 
larger absolute value of the work for smaller $\alpha$ and 
it is consistent with 
the fact that the Schwarzschild spacetime does not allow the quasistatic
deformation. 
The value of $(\delta W/m)/(\delta S/S)$ is proportional to $\alpha^{-4/3}$ 
in the $\alpha\gg 1$ region, 
although both $1-(\delta A/A)/(\delta S/S)$ and $3/2-(\delta m/m)/(\delta S/S)$
are proportional to $\alpha^{-2/3}$. 
This is probably because the terms which are proportional to
 $\alpha^{-2/3}$ in $\delta A/A$ and $\delta m/m$
cancel each other in the calculation of the work term, although we have not
proceeded this analysis. 
The work term rapidly decreases in the region $\alpha\gg 1$. 
For all $\alpha$, the absolute 
value of $(\delta W/m)/(\delta S/S)$ becomes large
as $l$ increases. The higher-multipole deformation requires
a larger work term.

\begin{figure}[tb]
\centering
{
\includegraphics[width=0.6\linewidth]{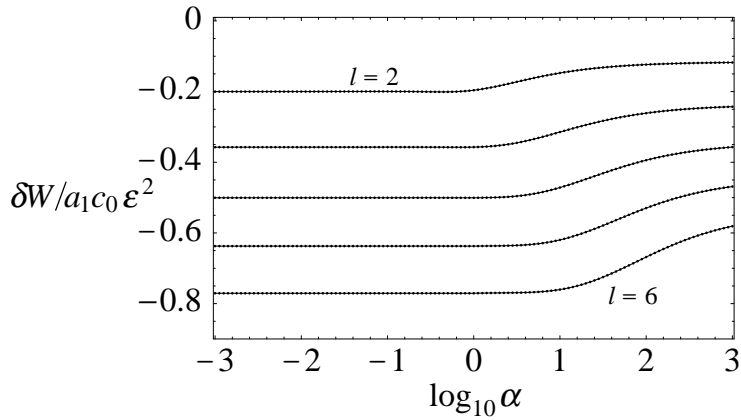}
}
\caption{The relation of $\alpha$ and the value of 
$\delta W/a_1c_0\epsilon^2$ for $l=2,...,6$.  
Here $\delta W/a_1c_0\epsilon^2$ is almost constant for $\alpha\lesssim 1$.
If we increase $\alpha$, $\delta W/a_1c_0\epsilon^2$ 
slightly changes and asymptotes to some nonzero constant value.
}
\end{figure}

Figure 5 shows the value of $\delta W/a_1c_0\epsilon^2$,
which was introduced with Eq.~\eqref{work_a1c0} 
in the approximate analysis for 
$\alpha\ll 1$ in Sec. IV.  
The value of $\delta W/a_1c_0\epsilon^2$ is almost constant in the region 
where the matching method provides a good approximation. 
Although this quantity 
slightly changes with an increase of $\alpha$, 
it seems to asymptote to some nonzero constant value.
This is supported by the following consideration. 
For $\alpha\gg 1$, we see that
$K^{(1)}\sim 1$ and $H^{(1)}\sim \alpha^{-2/3}/z$ 
from Eqs.~\eqref{H1_alpha_gg_1} and \eqref{K1_alpha_gg_1}.   
Using $z=r/r_h$ and $r_h\simeq 2m\alpha^{-2/3}$, we have 
$a_1\sim\alpha^{-4/3}m$ 
and thus $a_1c_0\epsilon^2\sim \alpha^{-4/3}m\epsilon^2$.
This leads to $-a_1c_0\epsilon^2\sim\delta W$ for $\alpha\gg 1$. 
Hence the value of $-\delta W/a_1c_0\epsilon^2$
is always $O(1)$ and the relation, Eq.~\eqref{work_a1c0},
approximately holds for all $\alpha$. 

We can construct another quantity which has the same order as $\delta W$
as follows:
\begin{equation}
-\frac{\delta W}{r_h}
\sim \left(1-\frac{r_h}{2m}\right)^{-1/2}
\left({c}_0^2-\tilde{c}_0^2\right). 
\label{work_deform}
\end{equation}
According to this formula, 
the work term normalized by the horizon radius depends only on two factors. 
One is related to the horizon radius, which shows that the 
absolute value of the work term becomes
larger for smaller $\alpha$. The other factor depends on
the values of $\tilde{c}_0$ and $c_0$, which are related to the deformation
of the horizon and spacelike infinity, respectively.  
Because the deformation of the two boundary surfaces 
becomes similar $|\tilde{c}_0/c_0|\simeq 1$ for larger $\alpha$,
the absolute value of the work term decreases with an increase of $\alpha$. 
Conversely, it becomes large with a decrease in $\alpha$,
which corresponds to a decrease in $|\tilde{c}_0/c_0|$.
Hence we interpret that the origin of the
work term is related to the difference of the deformation between 
the two boundary surfaces of the spacetime. 
The work would be necessary for the deformation
of both horizon and spacelike infinity, 
and their difference would appear in the first law as the work term.

\section{Summary and Discussion}

In this paper, we analyzed the static deformation of 
the Schwarzschild$-$anti-de Sitter black holes 
using perturbative techniques. We showed that 
there exists a regular solution for the first-order perturbation.  
The resulting spacetime contains the deformed black hole 
whose horizon deviates from the geometrically spherically symmetric surface.
The spacelike infinity of this spacetime is deformed simultaneously.
Hence this spacetime is not asymptotically anti$-$de Sitter, although
this is still weakly asymptotically anti$-$de Sitter in the sense of
the Ashtekar-Magnon definition~\cite{AM84}.  
In the $\alpha\ll 1$ case, the deformation of the horizon is about
$\alpha^l$ times smaller than the deformation of spacelike infinity,
while the deformations of the two are similar in the $\alpha\gg 1$ case.

We considered the $l=0$ mode of the second-order perturbation and
calculated the change in the horizon area and the mass. 
In the mass calculation, we used two definitions of the mass: 
the Abbott-Deser mass and the Ashtekar-Magnon mass. 
The Abbott-Deser mass for the deformed black hole diverges to minus infinity. 
If this result is realistic, we are forced to conclude that
the Schwarzschild$-$anti-de Sitter black hole would rapidly
deform toward the lower-energy state. 
But this definition is not gauge invariant 
at second order and we consider that this result may be spurious. 
Because the Ashtekar-Magnon mass is a covariant definition
of the mass from which we obtained finite results, 
we expect that it represents a real amount of energy although
one assumption is imposed in Eq.~\eqref{choosing_norm}
in choosing the norm of $\xi^t$.  
Our results indicate that the quasistatic deformation of the spacetime 
boundaries may occur with a finite change in the total energy 
in the weakly asymptotically anti$-$de Sitter spacetimes.  
Of course only with the analysis in this paper,
we cannot rigorously conclude that such a process actually occurs.   
But our expectation that the spacetime boundaries are flexible 
is also supported by the fact that 
no one has proved that spacelike infinity of the weakly asymptotically 
anti$-$de Sitter spacetime should be rigid; i.e., it 
has a global timelike conformal Killing vector field. 
There might exist many weakly asymptotically anti$-$de Sitter
solutions whose geometrical configuration at 
spacelike infinity temporally evolves.

We studied the thermodynamic 
first law of the deformed black holes using the Ashtekar-Magnon mass. 
In the $\alpha\gg 1$ case, the first law in the usual form is approximately
recovered, while in the $\alpha\ll 1$ case, the first law does not hold:
the contribution of the work term $\delta W$ becomes important. 
In Sec. V, we numerically calculated this work term in the range  
$10^{-3}\le\alpha\le 10^3$ for $l=2,...,6$ and confirmed that 
it is always negative in this regime. 
Let us discuss the implications of this first law from the viewpoint
of black hole thermodynamics. 
Although the ratio of $\delta m/m$ and $\delta A/A$ 
is fixed for each $\alpha$ in Secs. IV and V, 
we can consider these parameters to be independent 
if we further take account of the thermodynamical relation of the background 
Schwarzschild$-$anti-de Sitter spacetime, 
$\delta m_0=(\kappa/8\pi)\delta A_0$. 
The first law for the deformation process is given by 
$\delta m=(\kappa/8\pi)\delta A+\delta W$ with $\delta W<0$
for two independent parameters $\delta m$ and $\delta A$.
If the deformation process occurs without changing the horizon area 
(i.e., $\delta A=0$ and $\delta m=\delta W$), the black hole mass  
decreases (i.e., $\delta m<0$) and the black hole
evolves towards the lower-energy state.
Hence the deformation may work as a process of energy extraction from a 
spherical black hole. On the other hand, 
if there is a deformation process in which the black hole mass does not change,
the negative work term implies that
the horizon area of the deformed black hole increases in this process. 
This indicates that quasistatic deformation can be
a process consistent with the area theorem. 
Although whether the area theorem holds for these spacetimes 
is quite uncertain, we expect that this is an indication for the 
importance of the solution series of the deformed black holes.  
The value of $(\delta W/m)/(\delta S/S)$ 
is given in Fig. 4 as a function of $\alpha$. 
It is proportional to $\alpha^{-1}$ for $\alpha\ll 1$. 
This indicates that the deformation requires the larger absolute value of the  
work for smaller $\alpha$.
This is consistent with the fact that the Schwarzschild spacetime does not
allow a continuous static deformation. 
For $\alpha\gg 1$, the absolute value of 
$(\delta W/m)/(\delta S/S)$  
is proportional to $\alpha^{-4/3}$ and rapidly decreases with 
the increase in $\alpha$. The higher-multipole deformation
requires a larger absolute value of the work term. 
Because the work term satisfies 
the relation~\eqref{work_deform} for any $\alpha$,
this term is closely related to the difference of the deformation
between the two boundaries of this spacetime:
i.e., the horizon and spacelike two-surface at infinity.

Here, we discuss the reason why the work term is negative using
the Hartle-Hawking formula \cite{HH72} 
(see also \cite{Carter79} for a review). Using the Raychaudhuri equation, 
Hartle and Hawking derived the 
following formula for an increase in the horizon area in the quasistationary
evolution of a black hole:
\begin{equation}
\frac{\kappa}{8\pi}\delta A
=\oint dA\int_{t_0}^{t_1}\left(\frac{\sigma^2}{16\pi}+T_{ab}l^al^b\right)dt,
\label{Hartle-Hawking}
\end{equation}
where $t_0$ and $t_1$ denote the time for the initial state and final state,
respectively, $\sigma$ denotes the shear scalar of the null geodesic
congruence of the horizon, $T_{ab}$ is the energy-momentum tensor of the
matter field that crosses the horizon, and $l^a$ is the Killing vector field 
on the horizon. In the asymptotically flat case, this formula is equivalent to
the thermodynamic first law. If matter crosses the horizon and 
the flow of gravitational wave energy can be ignored, 
the second term of the integral in Eq.~\eqref{Hartle-Hawking}
becomes the change of mass $\delta M$ 
and the first term becomes zero because $\sigma$ has the
same order as $\delta A$ and thus $\sigma^2$ is 
much smaller than $\delta A$. 
(Here we do not consider any change of angular momentum.) 
If there is no matter and the gravitational wave 
energy is absorbed into the black hole, the first term has the same order
as $\delta A$ in this case and gives $\delta M$.
Hence the first law can be derived in general. 
The first term of the integral in Eq.~\eqref{Hartle-Hawking} 
has an analogy with the entropy generated by 
a surface shear viscosity with magnitude $\eta_\nu=1/16\pi$ of the ordinary
viscous fluid. Now we discuss the application of 
this Hartle-Hawking formula to the deformed black hole.
If some energy flux crosses the Schwarzschild$-$anti-de Sitter horizon
to induce a quasistatic deformation of the spacetime,
a part of the first term of the integral in Eq.~\eqref{Hartle-Hawking}
would contribute to the work term $\delta W$ necessary for the deformation,
while the remaining part (in addition to the second term) would contribute
to the mass term $\delta M$. 
The important fact 
is that the shear $\sigma$ is $O(\epsilon)$ in this case, and hence
all $\delta A$, $\delta M$, and $\sigma^2$ have 
the same order $O(\epsilon^2)$.  
The first term of the integral in Eq.~\eqref{Hartle-Hawking} cannot
be ignored, and introducing an unknown positive parameter $\beta$,
the work term and mass term may be given by 
\begin{align}
\delta W
&=-\oint dA\int_{t_0}^{t_1}(\beta{\sigma^2}/{16\pi})dt,\\
\delta M
&=\oint dA\int_{t_0}^{t_1}
\left(\frac{(1-\beta)\sigma^2}{16\pi}+T_{ab}l^al^b\right)dt,
\end{align}
which requires the work term to be negative. 
Although we have not investigated the validity of the relations
in terms of the Ashtekar-Magnon mass, 
this discussion provides a natural interpretation
of the reason why the work term 
obtained in this paper becomes negative.

Our remaining problems are as follows. As for the
problems of the solution sequence of deformed black holes, 
we should construct solutions beyond the perturbative region.
This would probably require numerical calculations.     
Because spacelike infinity is also deformed, we should impose the 
structure of spacelike infinity in the calculation. 
This arbitrariness in choosing the infinity structure
would lead to great degrees of freedom of the solution series 
of deformed black holes.  
The condition for the existence of a solution due to the choice
of infinity structure is of interest. 
Next, we should analyze the stability of spacetimes
with deformed black holes. This will require an analysis 
of the quasinormal frequency of the deformed black holes. 
At the same time, we would like to analyze the perturbation
from the Schwarzschild$-$anti-de Sitter black hole which gives 
the time-dependent geometrical configuration at spacelike infinity. 
With this analysis, 
what spacetime is an attractor of the weakly asymptotically anti$-$de Sitter 
spacetimes might become clear. 
Concerning the application of these spacetimes with deformed black holes,
these solutions might contribute to the brane world scenario. 
Although our analysis is restricted to four-dimensional cases,
it is natural to expect that similar solutions exist in higher dimensions.
Hence, by appropriate cutting and pasting, these spacetimes
might provide interesting models of the brane world scenario. 
Finally, we would like to analyze the implication for the AdS/CFT 
correspondence of these solutions. In the usual Schwarzschild$-$anti-de Sitter
black holes, the quasinormal frequencies have a relation to the
correlation function of the field on the boundary. 
By analyzing this correspondence in the spacetime with deformed black holes,
these spacetimes might shed new light on the AdS/CFT correspondence.
These are the problems we would like to investigate as the next step.

\acknowledgments

The work of H.Y. is supported in part by a grant-in-aid from 
Nagoya University 21st Century COE Program (ORIUM).



\begin{thebibliography}{99}


\bibitem{AGMOO00} O.~Aharony, S.~S.~Gubser, J.~Maldacena, H.~Ooguri, and Y.~Oz,
Phys.~Rep. {\bf 323}, 183 (2000).

\bibitem{RS99} L.~Randall and R.~Sundrum, 
Phys.~Rev.~Lett. {\bf 83}, 3370 (1999); {\bf 83}, 4690 (1999).

\bibitem{Lem95}J.~P.~S.~Lemos, Class.~Quantum~Grav. {\bf 12},
1081 (1995).

\bibitem{Lem95-2}J.~P.~S.~Lemos,
Phys.~Lett.~B \textbf{353}, 46 (1995). 

\bibitem{HL95}C.-g.~Huang and C.-b.~Liang, 
Phys.~Lett.~A \textbf{201}, 27 (1995).

\bibitem{LZ96} J. P. S. Lemos and V. T. Zanchin,
Phys. Rev. D {\bf 54}, 3840 (1996).

\bibitem{CZ96}R.~G.~Cai and Y.~Z.~Zhang,
Phys.~Rev.~D \textbf{54}, 4891 (1996). 

\bibitem{ABHP96} S.~\AA minneborg, I.~Bengtsson, S.~Holst, and P.~Peld\'an, 
Class.~Quantum~Grav. {\bf 13}, 2707 (1996).

\bibitem{Bri96} D.~R.~Brill, Helv.~Phys.~Acta {\bf 69}, 249 (1996).

\bibitem{Man97} R.~B.~Mann, Class.~Quantum~Grav. {\bf 14}, L109 (1997).

\bibitem{KMV98} D. Klemm, V. Moretti, and L. Vanzo,
Phys.~Rev.~D {\bf 57}, 6127 (1998);  
{\bf 60}, 109902(E) (1999). 

\bibitem{Hawk72}S.~W.~Hawking,
Commun.~Math.~Phys. {\bf 25}, 152 (1972). 

\bibitem{ACD02}M.~Anderson, P.~T.~Chru\'{s}ciel, and E.~Delay,
J. High Energy Phys. {\bf 10}, 063 (2002).

\bibitem{GSW03}G.~J.~Galloway, S.~Surya, and E.~Woolgar,
Class.~Quantum~Grav. {\bf 20}, 1635 (2003). 

\bibitem{GL93}R.~Gregory and R.~Laflamme,
Phys.~Rev.~Lett. {\bf 70}, 2837 (1993). 

\bibitem{Gub02} S.~S.~Gubser, Class.~Quantum~Grav. {\bf 19}, 4825 (2002).

\bibitem{Wise03} T.~Wiseman, Class.~Quantum~Grav. {\bf 20}, 1137 (2003).

\bibitem{KW03} H.~Kudoh and T.~Wiseman, 
Prog.~Theor.~Phys. {\bf 111}, 475 (2004).

\bibitem{RW57}T.~Regge and J.~A.~Wheeler, 
Phys.~Rev. {\bf 108}, {1063} (1957).

\bibitem{AM84}A.~Ashtekar and A.~Magnon, 
Class.~Quantum~Grav. {\bf 1}, L39 (1984).

\bibitem{CN02} P.~T.~Chru\'{s}ciel and G.~Nagy,
Adv.~Theor.~Math.~Phys. {\bf 5}, 697 (2002).

\bibitem{AD82}L.~F.~Abbott and S.~Deser,
Nucl.~Phys. {\bf B195}, 76 (1982).

\bibitem{Van97}L.~Vanzo, 
Phys.~Rev.~D \textbf{56}, 6475 (1997).

\bibitem{CCK00}
M.~M.~Caldarelli, G.~Cognola, and D.~Klemm,
Class.~Quantum~Grav. {\bf 17}, 399 (2000).


\bibitem{Sil02}
S.~Silva, Class.~Quantum~Grav. {\bf 19}, 3947 (2002).

\bibitem{Bar03}
G.~Barnich, Class.~Quantum~Grav. {\bf 20}, 3685 (2003).

\bibitem{GNPP00} R.~J.~Gleiser, C.~O.~Nicasio, R.~H.~Price, and J.~Pullin,
Phys.~Rep. {\bf 325}, 41 (2000).

\bibitem{Kan96}G.~Kang, 
Phys.~Rev.~D {\bf 54}, 7483 (1996).

\bibitem{HH72}J.~B.~Hartle and S.~W.~Hawking, 
Commun.~Math.~Phys. {\bf 27}, 283 (1972).

\bibitem{Carter79}B.~Carter, ~in {\it General Relativity},
edited by S.~W.~Hawking and W.~Israel (Cambridge University Press,
Chambridge, England, 1979).

\end{thebibliography}
\end{document}